# Temperature behavior of NaI (Tl) scintillation detectors

K.D.Ianakiev[11], B. S. Alexandrov[1], P.B.Littlewood[2], and M.C.Browne[1]

[1]**Nuclear Nonproliferation Division, Los Alamos National Laboratory,**

**Los Alamos, NM 87545, USA**

[2] **Cavendish Laboratory, Cambridge University, Cambridge, UK**

**ABSTRACT**

**It is a familiar fact that the total measured light yield of NaI (Tl) detectors is a nonlinear function of temperature. Here we present new experimental data for the temperature behavior of doped NaI(Tl) scintillators that instead shows a *linear* dependence of light output over a wide temperature range— including that for outdoor applications. The shape of the light pulse shows in general two decay processes: a single dominant process above room temperature and two decay time constants below. We show that redistribution of the intensities is temperature-dependent; the second (slow) decay component is negligible at room temperatures, but, by –20°C, it contributes up to 40% of the total light and has a duration of several microseconds. We discuss the profound effect this new understanding of the light output has on the pulse height analysis instrumentation. We introduce a theoretical model to explain the experimental results. In addition, we describe a unique technique for correcting both amplitude and shape temperature changes inside the NaI(Tl) detector package.**

## 1. INTRODUCTION

The NaI(Tl) scintillator is one of the first gamma detectors used for spectroscopy. Even now, many decades after its establishment, it continues to be the first choice for outdoor gamma spectroscopy because of its combination of cost efficiency and reliability. The very complicated process of converting the ionizing energy into an electrical signal causes a complex temperature dependence of the scintillator's light output. Its properties were intensively studied in the 1960s and described in major textbooks and monographs.[1,2]

It is common practice to consider the luminescent response of NaI(Tl) scintillators as a single dominant process with a temperature-dependent decay constant and complex temperature dependence of the fluorescent yield. Typical data for the light yield temperature dependence are the famous data from the Harshaw catalog[3] and more recent data.[4] The classic understanding of the temperature behavior comes from the three-level theoretical model,[2,5] which explains the temperature dependence as a consequence of nonradiative transitions between activator levels.

The experimental data for temperature dependence of the light decay time constant, presented by different authors[2,4,6], have been fitted to a single component, whose temperature dependence becomes very steep below room temperature. This is not with agreement with the theoretical model, which predicts a smooth change of the time constant over the absolute temperature range. The disagreement between the experimental data and the model for the scintillator light yield is even greater: the model calls for weak monotonic exponential temperature dependence; yet the experimental data for all doped halide scintillators shows a broad maximum at room temperature and a steep decline of the light yield below 0°C. Furthermore, the experimental data from different authors has been inconsistent. To explain the observed nonlinear behaviour of the light output for NaI (Tl), an additional function *f(Temperature)* that represents probability of occupation of the activation centres has been introduced.[2]

The developers of the original detector technology in fact observed a secondary light component[2] with a longer time constant. The time dependence of NaI(Tl) fluorescence was measured precisely at room temperature with the delayed coincidences method,[7] and two dominant decay time components were observed. Because 95% of the total light yield is collected within the first 800 ns, the influence of the second component has been neglected, and use of shapers with about 1 μs peaking time has been recommended. It is our belief that this recommendation and the desire for higher throughput made a short shaping time (1 μs and below) the common choice for NaI(Tl) detectors.

In recent years, the awareness of the impact of signal processing on the temperature drift has increased.[8] Those measurements,[8] using a very long shaping time, in fact showed a nearly linear T-dependence of the light output, although this attracted no comment, and the physical reasons were not discussed.

The light output of another typical two-component halide scintillator—CsI (Tl)—was extensively studied experimentally and theoretically in the past[9] and currently in the light of its use with a photodiode light sensor.[10] The relative intensities and time constants of both fast and slow components were measured over a wide temperature range, and an explanation of the origin of the two components was given.[11]

One of us observed the slow component at room temperatures long ago and found it to be a limiting factor for high count rate applications (e.g., for the spent fuel attribute test) using NaI(Tl) detectors.[12] A unique preamplifier circuitry with rise time equal to the rise time of the photomultiplier tube (PMT) current pulse and single exponential decay time constant were developed to allow operation at input count rates up to $10^6$ cps.[13] Later, the temperature dependence of the light yield of NaI(Tl) scintillators was measured in 1998 as a corollary of work to install a scintillation-detector monitoring system in BN-350 breeder reactor, at Aktau, Kazakhstan. The first public disclosure of the results was a patent application filed[14] because of increased interest in that problem after the events of 9/11.

In this paper, we present comprehensive experimental data and an outline of a quantitative model explaining the temperature behavior of doped NaI(Tl) scintillators. We outline technical solutions for overcoming the deleterious effects of that behavior. Our study shows linear T-dependence of the NaI(Tl) light output over a wide temperature range including that for outdoor applications. The normalized shape of the light pulse shows a dual character:

*i)* A single dominant rate constant for the fluorescent decay above room temperature.

*ii)* The appearance of a second (slow) component below room temperature, with a temperature-dependent redistribution of the intensities.

The second decay component is negligible at room temperatures, but, at –20°C, it occupies up to 40% of the total light and lasts several microseconds.

We believe that not taking the correct time profile of the light pulse into account has been misleading users and developers over the decades. The artificial broad maximum in the light yield comes from a convolution between the finite width of the shaper response function and the temperature-dependent shape of the light pulse. That convolution causes profound dependence on all characteristics of NaI(Tl) spectrometers.

## 2. EXPERIMENTAL RESULTS

Considering the temperature dependence of the duration of NaI(Tl)[6] light pulse and shaping with a peak time shorter than the total light pulse led us to the idea that the previously reported[3] nonlinear temperature dependence does not describe the total light output. Here we prove that this temperature dependence is an artificial result of convolution between the light pulse and shaper response if changed with the temperature duration of PMT current pulse and finite duration of shaper response. So the shaper will not "see" an increasing fraction of the light pulse when the temperature is reduced below 0°C. This effect is the opposite of the negative temperature coefficient of the light yield caused by nonradiative transitions and thus will cause a temperature dependence with a broad maximum.

A shaper with a flat top equal to or longer than the duration of the light pulse will eliminate effects resulting from changes in the temporal dependence of the light emission. The Gated Integrator (GI) is the simplest and most convenient shaper that can provide a signal proportional to the integrated amount of light emitted. To demonstrate the effect of shaping, we have compared the Na(Tl) spectrum obtained with semigaussian and Gated Integrator shapers using a broad range of shaping time constants. In a simple experiment, we placed the whole PMT and scintillator assembly in an environmental chamber with the electronics at ambient temperature. Thus, the measured temperature effect is the result of temperature dependences of both the scintillation crystal and the PMT. Because the temperature coefficient of the PMT is almost constant over a wide temperature range,[21]it will add to the temperature dependence arising from the crystal scintillation. Figure 1 shows the block diagram of the experimental setup.

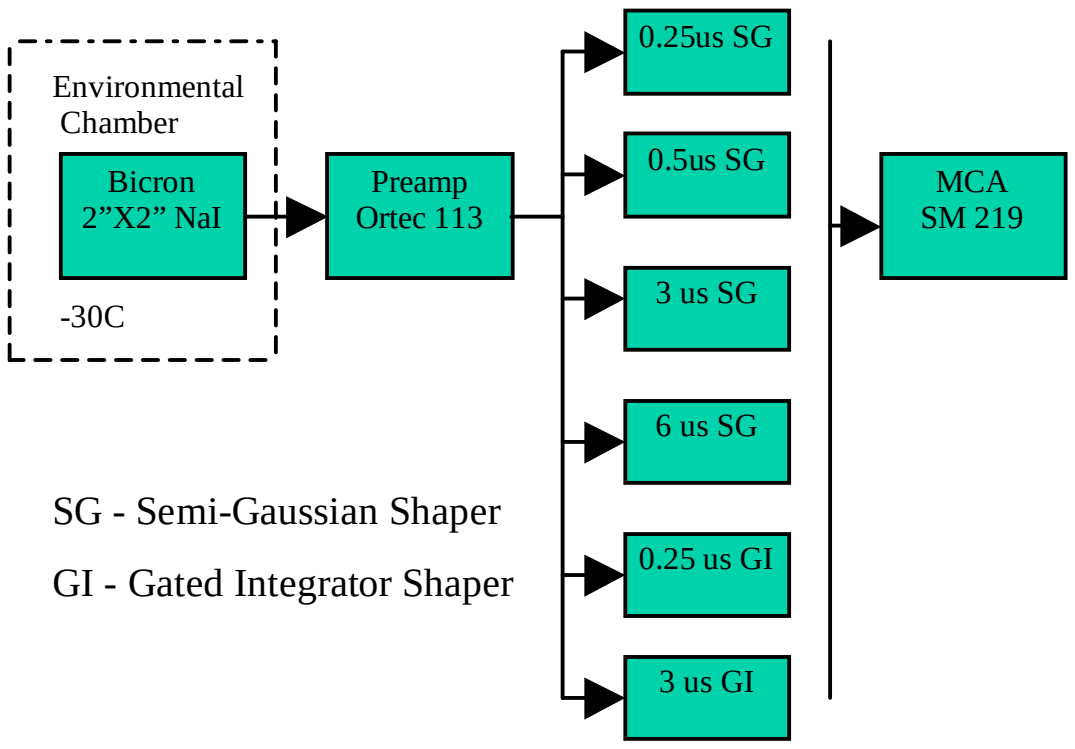

***Fig. 1. Block diagram of the experimental setup for measuring temperature dependence of NaI(Tl) scintillator light output with different shapers.***

We placed a standard 2-in. × 2-in. NaI(Tl) detector Bicron model number_____ .in the environmental chamber. Standard network interface module (NIM) electronics have been used for the shapers and multichannel analyzer (MCA) with exception of the following modification that provides accurate integration of PMT pulses with microsecond duration:

- Increase the time constant of the anode and preamplifier decoupling network above 50 ms (the typical value provided by most scintillation detectors' manufacturers is a couple of hundred microseconds).
- Disable the gate time out circuitry of the gated integrator to integrate the whole prefilter pulse.

The chamber temperature changed at a rate of 3°C/hour, and each set of measurements at a given temperature was taken after 8 hours' hold time to allow excellent thermal equilibration. We digitized the PMT current pulses with a Tektronix digital scope for each temperature. The response corresponds to excitation around 662 keV. To study the effect on pulse shape alone, we normalized all pulses to unit area, as shown in Fig. 2. The height spectra for each shaper and temperature have been recorded and the 662-keV peak position and the full width at half maximum (FWHM) calculated. The temperature dependence of the light pulse can be explained by redistribution between prompt and delayed components (as measured in [7]), rather than by a single temperature-dependent rate constant. That redistribution is not uniform: the slow component is negligible above room temperatures and increases for lower temperatures. Even though the amplitude of the PMT current in the slow component is a negligible fraction of fast component, the integral contains a substantial part of the total light emission because of its duration. At a temperature of –20°C, the slow component contains 40% of the total light and lasts many microseconds. To explain the effect of changed temporal shape on overall detector pulse height characteristics, we have simulated pulse processing of normalized pulses in Fig. 3 for a 0.5 µs semigaussian shaper and a GI.

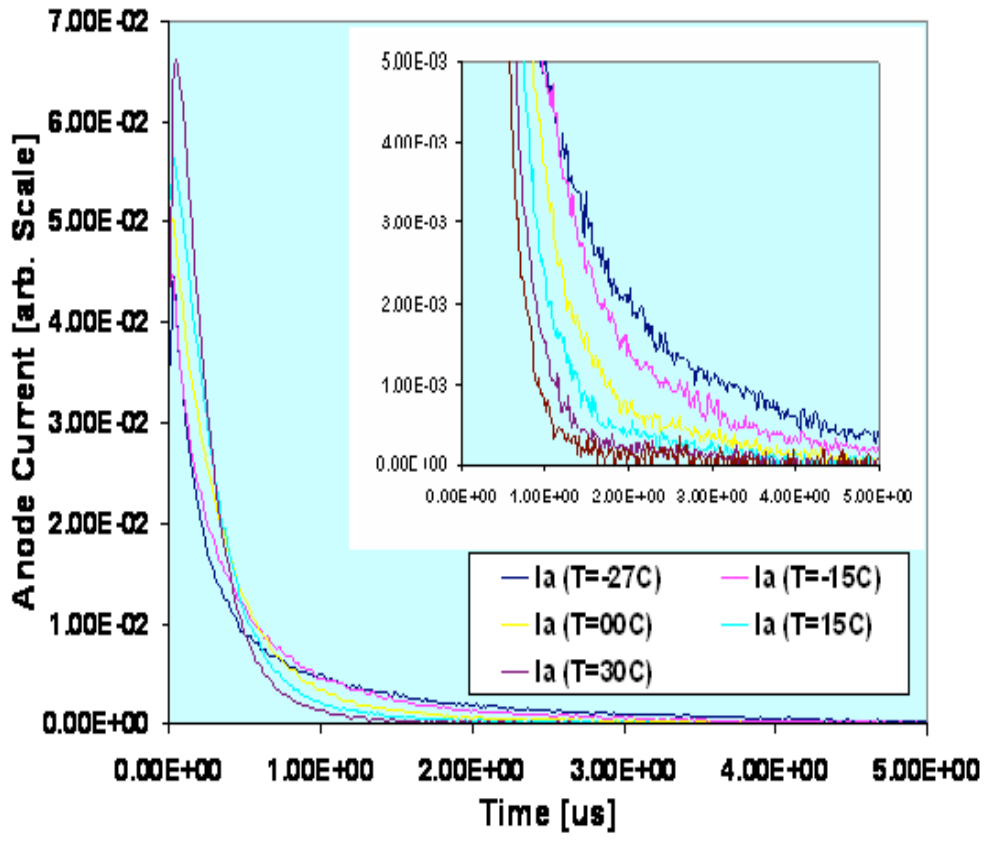

***Fig. 2. Normalized to the total area of PMT anode current pulses for different ambient temperatures. A temperature-dependent redistribution between a fast and a slow component is evident. In the inset, we have expanded the vertical scale.***

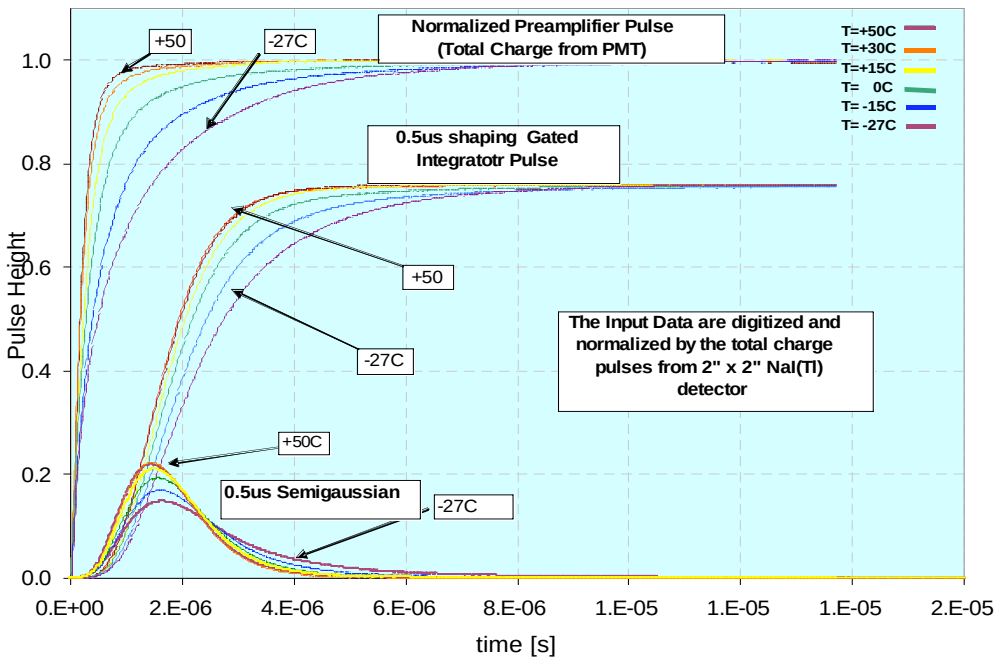

***Fig. 3. Temperature effect on pulse height spectra is demonstrated by measuring the relative change of the $^{137}$Cs peak position with different shapers and different shaping methods. Shown are simulated response of the preamplifier (upper), gated integrator (middle), and 0.5 µs shaper (lower) to the normalized current pulses of Figure 2.***

The relative change of the peak position with temperature (normalized to +50°C) is shown in Fig. 4.

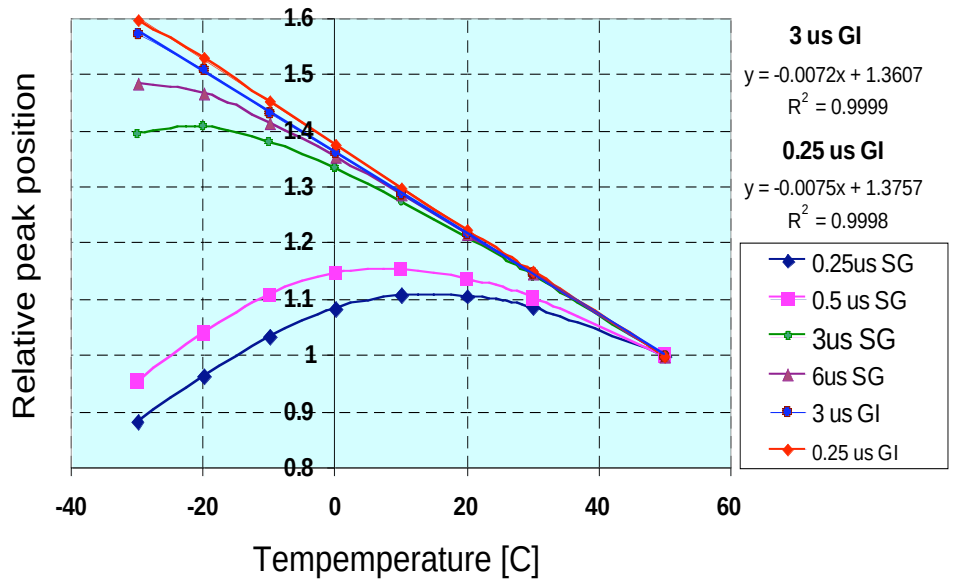

***Fig. 4. Temperature dependence of NaI(Tl) pulse height spectrum using different shaping and time constants. All the data is normalized to a +50°C 662-keV peak position versus ambient temperature from different shapers.***

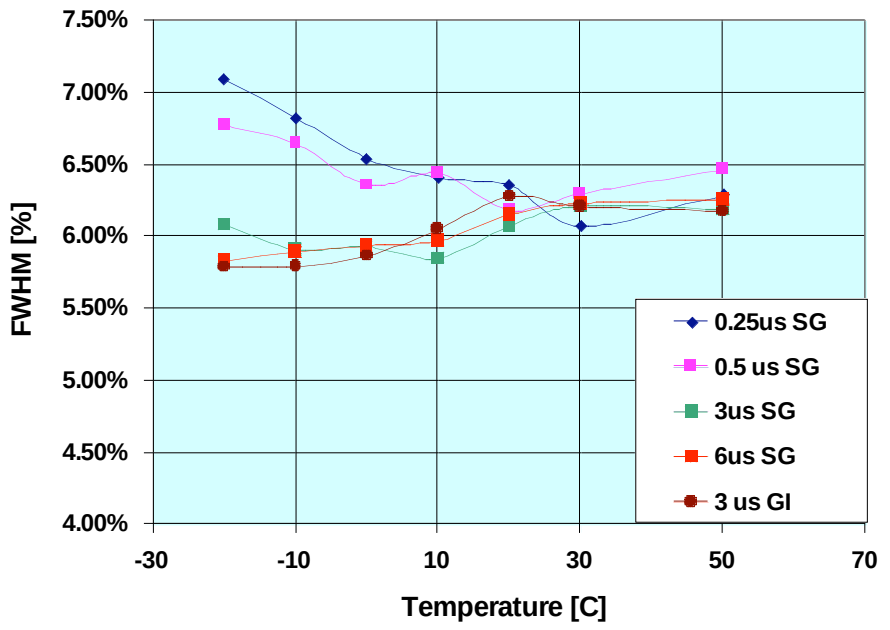

***Fig. 5. Temperature dependence of $^{137}$Cs 662-keV peak FWHM for different shapers.***

The data in Fig. 4 shows a profound dependence on the shaper used. The peak shift for the 0.25 µs semigaussian shaper has the closest shape to the widespread Harshaw data[3] with a maximum around +15°C. With increasing shaping time, the maximum is shifted toward negative temperatures, and the temperature characteristic becomes more linear. The gated integrator linearized the characteristic completely—independent of the time constant or Gaussian prefilter.

The temperature does not affect the FWHM as badly as the peak position. The resolution does not depend on the shaping time constant above room temperature because all shapers "see" the total light. But below room temperature where the actual redistribution between the light components occurs, the difference is observable. That fact does not have significant practical consequence except to confirm the dual temperature behavior of Na(Tl) light output.

## 3. THEORETICAL MODEL

Our model for the temperature-dependent behavior of the NaI (Tl) scintillators is based on two dominant components of the light pulse we measured: one fast with a time constant of 230 nsec and one slow with a time constant of about 1 µsec. These two components correspond to two different processes in the crystal. The physical picture of the processes briefly is as follows:

In the very beginning of the process, we have production of electron-hole pairs along the path of the ionizating particle. Because of the very strong coupling between the electron and holes in alkali halide crystals,[15] excitons are formed in a mobile exciton band followed by rapid conversion into Self-Trapped Excitons (STE) because of the interaction with the ions of the crystal lattice.[16] The fast component of light emission results from the prompt capture of the STE by the ($Tl^+$) level.[17]

Note that all the light emitted (from both slow and fast processes) comes from the *same* optical transition on the activator level. In our model, the redistribution of the light output arises from a competition between different mechanisms for reaching the activator levels. This redistribution occurs by temperature dependence of the population of the STE.[18] The temperature dependence exists because the STE can reach the (Tl$^+$) level by thermal-assisted hopping[16] followed by capture and recombination. This process competes with the multi-phonon-assisted dissociation of the STE[19] and transport to the activator level by independent electrons and holes. In this way, a part of the STE can "feed" the population of excitons responsible for the fast component of the light output at every temperature. The rest of the STE, through multi-phonon-assisted dissociation, dissociate into electrons and holes. After this multi-phonon-assisted dissociation, the separate carriers by binary diffusion can reach and be captured by the (Tl$^o$) center and then recombine.[17] This second mechanism is responsible for the amplitude of the slower component of the light output of the NaI (Tl) scintillators. Figure 5 shows a schematic of the model, which gives rise to a set of coupled rate equations. A more detailed analysis of this model will appear elsewhere.[20]

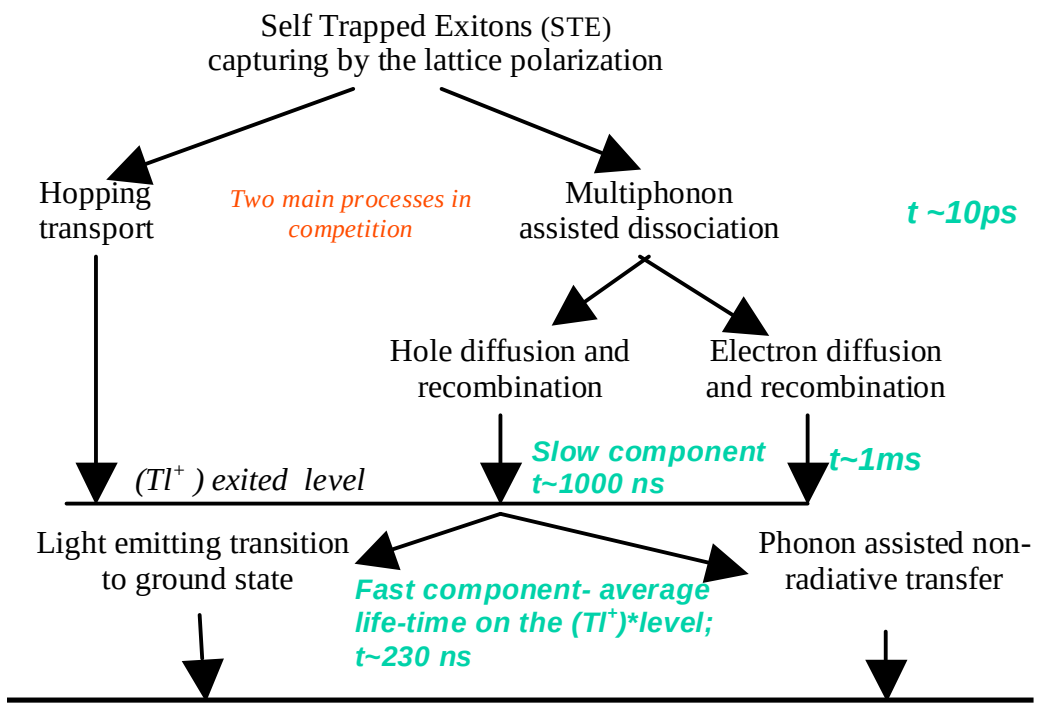

***Fig. 6. Scintillation mechanism arising from two pathways for reaching the activator's centers.***

## 4. SIGNAL PROCESSING OF NaI(Tl) LIGHT PULSES

We have shown that the temperature changes the light pulse shape and duration in a very wide range from one microsecond to more than 10 microseconds at negative temperatures. The variable shape of the signal is much more difficult to handle than the changed amplitude. Therefore, we have developed a proprietary technology for the correction of temperature changes of both amplitude and shape of PMT current pulses.[14] It is based on the unique property that the weighted sum of a prompt and integrated exponential function is a step function. The solution for a single exponential component is rather intuitive: a resistor connected in series with the integrating capacitor. The output signal is a sum of the voltages across the R-C network:

$$U_{out} = \frac{Q}{C}\left[1 - \exp\left(-\frac{t}{\tau}\right)\right] + R\frac{Q}{\tau}\exp\left(-\frac{t}{\tau}\right)$$

when $R = \tau / C$, then $U_{out} = \frac{Q}{C} 1(t)$ .

A similar scheme can be applied to two exponential components' light signal (Fig. 7), where the feedback circuitry consists of two parallel R-C networks.

$$I_{sc} = \frac{Q_1}{\tau_1}\exp\left(-\frac{t}{\tau_1}\right) + \frac{Q_2}{\tau_2}\exp\left(-\frac{t}{\tau_2}\right)$$

The second network C-R2 is connected through a voltage divider providing fraction ***m*** of output voltage. Note that a similar technique has been used to simulate the light response of CsI(Tl) for calibration purposes.[22]

The Laplacian transform of the PMT current pulse $I_{sc}(s)$ can be presented as

$$I_{sc}(s) = \frac{Q_1}{\tau_1}\left(\frac{\tau_1}{1 + s\tau_1}\right) + \frac{Q_2}{\tau_2}\left(\frac{\tau_2}{1 + s\tau_2}\right),$$

where Q1 and Q2 are the charges corresponding to two exponential components; $\tau_1$, $\tau_2$ are the time constants of the fast and slow components.

The configuration of Fig. 7 provides a step response to the two-exponential-component PMT current pulse.

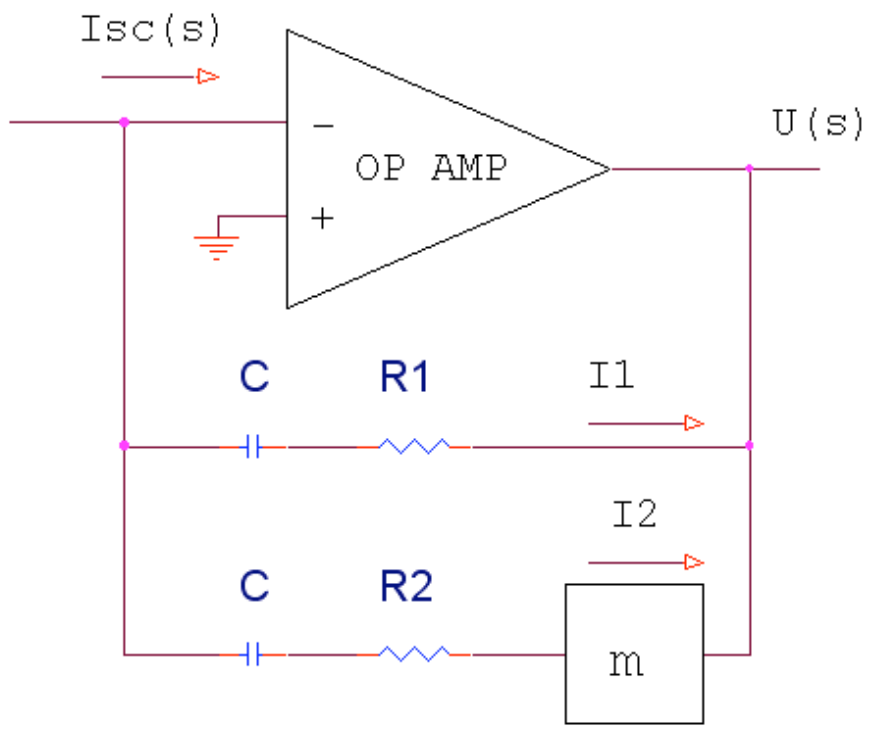

***Fig. 7. Step response preamplifier for two-exponential-component PMT pulse ($Q_1$,$\tau_1$ $Q_2$, $\tau_2$).***

The Laplace transform of the preamplifier output voltage can be written as

$$U(s)=\frac{Q_1+Q_2}{sC(1+m)}\cdot\frac{\dfrac{1+s\cdot\dfrac{Q_1\tau_2+Q_2\tau_1}{Q_1+Q_2}}{(1+s\tau_1)(1+s\tau_2)}}{\dfrac{1+s\cdot\dfrac{R_2C+mR_1C}{1+m}}{(1+sR_1C)(1+sR_2C)}} \qquad /1/$$

By selecting the divider value m proportional to the ratio of the light components $Q_1/Q_2$ with RC values equal to the time constants $\tau_1$, $\tau_2$, we have

$$U(s)=\frac{Q_1+Q_2}{sC(1+m)} \quad \text{or} \quad U(t)=\frac{Q_1+Q_2}{C}\cdot 1(t)\cdot$$

The technique described above can be used for compensation of both time constants or only one of them. By compensating only the slow component ($R_1$=0), the shape of the signal is constant with rise time less than one microsecond over the temperature range,

$$U(s)=\frac{Q_1+Q_2}{sC(1+m)}\cdot\frac{1}{(1+s\tau_1)}\ .$$

This allows use of any MCA and implementation of simple temperature correction in the detector package.[14] The proposed[18] method for stabilization allows use of a very low intensity $^{40}$K background source for peak tracking and prompt compensation for any rough or sudden temperature drift.

The exact value of the second component can be measured by using only the first R-C network R1, C1, thus compensating the fast component only. This scheme is applicable also for pulse-shape analysis in other two-component scintillators for gamma-neutron separation.

## 5. CONCLUSIONS

We have observed the temperature dependence of NaI(Tl) light output as a temperature-dependent redistribution between fast and slow components. The slow component is negligible at room temperature, but it accommodates a substantial fraction of the total light output at temperatures below 0°C. We measured a linear temperature dependence of the fluorescent yield of NaI(Tl) scintillators in the outdoor temperature range when the total light pulse is integrated. The broad maximum in the previously published data is thus explained as a convolution between the temperature-dependent shape of the light pulse and the shaper response. Our theoretical model explains the above behavior as arising from two competing processes (exciton hopping and binary diffusion) to reach the activation centers.

A proprietary technique that compensates for the temperature dependence of the light pulse has been developed. It provides a step response or single exponential rise time of the preamplifier output, thus compensating for the redistribution between the slow and fast components. This technique will improve the high count rate performance of NaI(Tl) scintillation detectors and mitigate the temperature effect on the shape of light pulse.

## References:

1. Glenn F. Knoll, *Radiation Detection and Measurement*, John Wiley & Sons (1999).
2. J. Birks, *The Theory and Practice of Scintillation Counting*, Pergamon Press (1964).
3. *The Harshaw Catalog of Optical Crystals*, (Harshaw Chemical. Company, Solon, Ohio, 1967).
4. C.L. Melcher, *IEEE Transactions on Nuclear Science*, Vol. 35, No 1 (1988).
5. H. Jones and NF Mott, *Proc. Roy. Soc.* A162, 49 (1937).
6. J.S. Schweitzer, et al., *IEEE Transaction on Nuclear Science*, Vol. NS-30, No1 1983.
7. L. M. Bollinger and G. E. Thomas, *The Review of Scientific Instruments*, Vol. 32, (1961).
8. C. Rozsa, C. Grodsinsky, D. Penn, P. Raby, R. Schreiner, Bicron, Newbury Nuclear Science Symposium, 1999. Conference Record. 1999 IEEE.
9. R. Gwin and R.B. Murray, *Phys. Rev.*, vol. 131, p 501 (1963).
10. R. B. Murray and A. Meyer, *Phys. Rev.*, vol. 122, p 815 (1961).
11. J. Valentine et al., *Nucl. Instr. And Methods in Phys. Research* A325 (1993) 147-157C.L.
12. R. B. Murray, *IEEE Transaction on Nuclear Science*, Vol. NS-22 (1975).
13. R. Arlt, personal communication.
14. K. Ianakeiv, T. Grigorov, IAEA consultant agreement.
15. K. Ianakiev, M.C. Browne, J. Audia, W. Hsue, Patent application S102,322 (2004).
16. W.J. Van Sciever, 1955 High Energy Physics Rpt, No 38 Stanford University (1955).
17. H. Nishimura and S. Nagata, J*ournal of Luminescence* 40&41, 429-430, (1988).
18. H. B. Dietrich , A. E. Purdue, R. B. Murray, and R.T. Williams, *Phys. Rev. B*, vol. 8, p. 5894 (1973).
19. B. S. Alexandrov, J. Audia, M. C. Browne, K. Ianakiev, H. Nguyen, P. Reass, R. Parker, Proceedings of the INMM (2005).
20. B. Goodman and O.S. Oen, *J. Phys. Chem. Solids* Vol. 8, p. 291 –294 (1959).
21. B. S. Alexandrov, K. Ianakiev, P. Littlewood, in preparation.
22. *Photomultiplier Tubes Principles and Applications*, Photonis, Brive, France.
23. J. Valentine, V. Jordanov, D. Wehe, G. Knoll, NIM A314 (1992).